\documentclass[aps,prl,twocolumn,groupedaddress]{revtex4}

\usepackage{graphicx} 

\begin{document}

\renewcommand{\thefootnote}{\alph{footnote}}

\preprint{APL, Y. Takahashi}

\title{
Room-temperature excitonic absorption in quantum wires -submitting to APL- 
}

\author{
Yasushi Takahashi\footnote[1]{Electronic mail: taka8484@issp.u-tokyo.ac.jp}, 
Yuhei Hayamizu, 
Hirotake Itoh, 
Masahiro Yoshita\footnote[2]{Also a visiting scientist at Bell laboratories, Lucent Technologies}, 
and Hidefumi Akiyama$^\dagger$}

\affiliation{
Institute for Solid State Physics, University of Tokyo, and CREST, JST,\\
5-1-5 Kashiwanoha, Kashiwa, Chiba 277-8581, Japan
}

\author{Loren N. Pfeiffer and Ken W. West}

\affiliation{Bell Laboratories, Lucent Technologies, 600 Mountain Avenue, Murray Hill, NJ 07974}

%
%
\date{June 10, 2005}

\begin{abstract}
We measured the absorption spectra of T-shaped quantum wires at room temperature using waveguide-transmission spectroscopy. A strong and narrow room-temperature one-dimensional-exciton absorption peak was observed, which indicates a peak modal absorption coefficient of 160 cm$^{-1}$ per 20 wires with a $\Gamma$-factor of $4.3\times10^{-3}$, a width of 7.2 meV, and strong polarization anisotropy.
\end{abstract}

\maketitle

\newpage

The optical properties of quantum wires have been intensively studied toward achieving novel device applications \cite{arakawa1982,asada1985}. So far, most research has relied on emission measurements such as photoluminescence (PL) \cite{akiyamaPL}, PL excitation (PLE) \cite{akiyamaPL,itoh2003}, and lasing \cite{kapon1989,weg1993,hayamizu2002,yagi2003}, and there have been no measurements of quantitative direct absorption on quantum wires. We recently measured the absorption spectrum of a single T-shaped quantum wire (T-wire) embedded in an optical waveguide using straightforward waveguide-transmission spectroscopy at 5 K, and found that the one-dimensional (1D)-exciton ground state has a large modal absorption coefficient of 80 cm$^{-1}$ despite the small lateral size of the T-wire, i.e., 14 nm $\times$6 nm \cite{takahashi2005}. The exciton peak exhibited thermal broadening at high temperatures, and room-temperature absorption was not measurable. 

It is very important to measure room-temperature quantitative absorption in investigation of the application of quantum wires to practical optical devices. However, the absorption spectra of quantum wires at room temperature have not been measured even with PLE.

\begin{figure}[bp]
\centerline{\scalebox{1.0}{\includegraphics{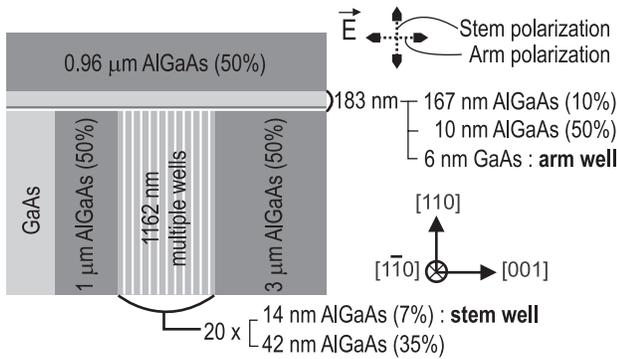}}}
\caption{Schematic view of 20-wire device, where 20 T-wires are embedded in T-waveguide. Cross-section of each T-wire is 14 nm (stem well)$\times$6 nm (arm well). Percentages in parentheses represent Al content $x$ of $\rm{Al_{x}Ga_{1-x}As}$.}
\label{fig1}
\end{figure}

In this letter, we report the absorption spectra of 20 T-wires embedded in an optical waveguide at 297 K and 5K using waveguide-transmission spectroscopy. Because of the increased overlap between the wires and the optical waveguide compared with the previous single-wire device, the 20-wire device has a strong absorption peak at room temperature, demonstrating room-temperature 1D-exciton absorption. The 1D-exciton absorption peak has a maximum value of 160 cm$^{-1}$ and a full width at half maximum (FWHM) of 7.2 meV. In addition, the absorption by the T-wires has strong polarization anisotropy. 

\begin{table}[bp]
\caption{Comparison of structures and measured data for 20-wire and single-wire devices.  Listed absorption data are for arm polarization. Peak area is evaluated as product of peak value and FWHM. Ratio is evaluated for values $(a)$ and $(b)$.}
\begin{center}
\begin{tabular}{l|cc|c|c}
\hline
sample & \multicolumn{2}{c|}{20 wire} & single wire & ratio \\
\hline
T-wire size (nm$^2$) & \multicolumn{2}{c|}{14$\times$6} & 14$\times$6 & \\
Waveguide size (nm$^2$) & \multicolumn{2}{c|}{1162$\times$183} & 514$\times$127 & \\
confinement factor $\Gamma$ & \multicolumn{2}{c|}{4.3$\times$10$^{-3}$ $^{(a)}$} & 4.6$\times$10$^{-4}$ $^{(b)}$ & 9.3 \\
\hline
temperature (K) & 297 & 5 & 5 & \\
coupling efficiency $\eta$ & 0.40 & 0.40 & 0.24 & \\ 
continuum $\alpha$ (cm$^{-1}$) & & 150 $^{(a)}$ & 16 $^{(b)}$& 9.4 \\
exciton peak $\alpha$ (cm$^{-1}$) & 160 & & 80 & \\
FWHM (meV) & 7.2 & $<$2 & 1.6 & \\
peak area (cm$^{-1}$meV)  & 1152 $^{(a)}$ & & 128 $^{(b)}$ & 9.0 \\
peak energy (eV) & 1.4884 & 1.5808 & & \\
\hline
\end{tabular}
\label{table1}
\end{center}
\end{table}

Figure \ref{fig1} is a schematic view of a 20-wire device fabricated by the cleaved-edge overgrowth method with molecular-beam epitaxy\cite{pfeiffer1990} and a growth-interrupt annealing technique\cite{yoshita2002}. The 20 T-wires are formed at the T-shaped intersections of 20 multiple (001) $\rm{Al_{0.07}Ga_{0.93}As}$ quantum wells (stem wells) and a (110) GaAs quantum well (arm well). The T-wires are embedded in the core of the T-shaped optical waveguide (T-waveguide) with a lateral size of 1162 nm$\times$183 nm, surrounded by $\rm{Al_{0.5}Ga_{0.5}As}$ cladding layers. The cavity length $L$ is 512 $\mu$m, and cavity facets are uncoated. Details on sample fabrication and characterizations are discussed in separate papers \cite{ Akiyama2003,itoh2003,takahashi2003}. 

The upper columns in Table \ref{table1} summarize the structural parameters of the 20-wire device, in comparison with those of the single-wire device used in the previous low-temperature transmission study \cite{takahashi2005}. Each T-wire has the same structure of 14 nm (stem well)$\times$6 nm (arm well) formed with the same materials in both devices, while the T-waveguide structure is changed to optimize optical confinement factor $\Gamma$ for the lowest waveguide mode. We calculated $\Gamma$ using a finite element method and obtained $\Gamma=4.3\times10^{-3}$ for the 20-wire device. This value is larger than that for the single-wire device by a factor of 9.3.

The experimental method for waveguide-transmission spectroscopy was described previously\cite{takahashi2005}. In the present experiment, we measured transmission spectra at 297K and 5K for two linear polarizations parallel to the arm well (denoted as arm polarization) and the stem well (stem polarization). Coupling efficiency $\eta$ of incident light to the T-waveguide was 0.4 at both temperatures. 

\begin{figure}[tp]
\centerline{\scalebox{1.0}{\includegraphics{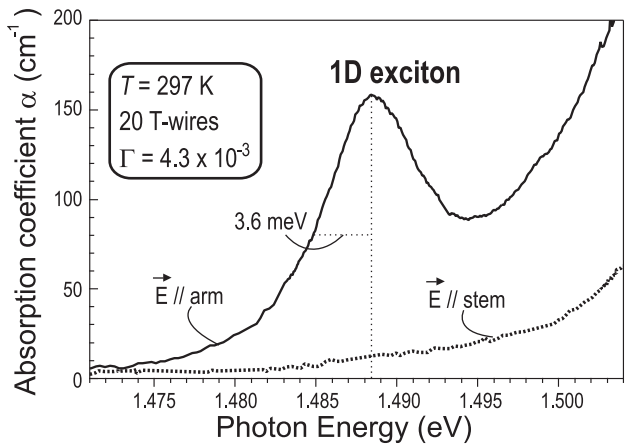}}}
\caption{Absorption spectra of 20 wires at 297 K. Solid curve and dotted curve represent absorption spectra for arm and stem polarizations, respectively. Room-temperature 1D excitonic absorption is observed at 1.4884 eV.}
\label{fig2}
\end{figure}

Figure \ref{fig2} plots the absorption spectra for 20 wires at 297 K. The longitudinal axis represents modal absorption coefficient $\alpha$. The solid curve represents the absorption for arm polarization, and the dotted for stem polarization. As will be explained later, the absorption peak at 1.4884 eV for arm polarization is due to the 1D-exciton ground state.

The absolute value of the absorption coefficient, $\alpha$=160 cm$^{-1}$, for the 1D exciton peak is one of the most important results in this paper. The value of $\alpha$=160 cm$^{-1}$ gives a very small transmission probability of e$^{-\alpha L}$=2.8$\times 10^{-4}$ when $L$ is 512 $\mu$m. This demonstrates that the quantum wires in an optical waveguide have strong excitonic absorption for light propagating along the wires even at room temperature, despite the small optical confinement factor $\Gamma$, which is about one order of magnitude smaller than those of ordinary multiple-quantum-well devices. 

The 1D-exciton peak at 1.4884 eV is indicated by the dotted vertical line in Fig. \ref{fig2}. This peak is overlapped by the absorption tail of the arm well at higher energy. Thus, the FWHM of the peak is estimated as 7.2 meV from half width 3.6 meV of half maximum on the lower energy side. 

\begin{figure}[tp]
\centerline{\scalebox{1.0}{\includegraphics{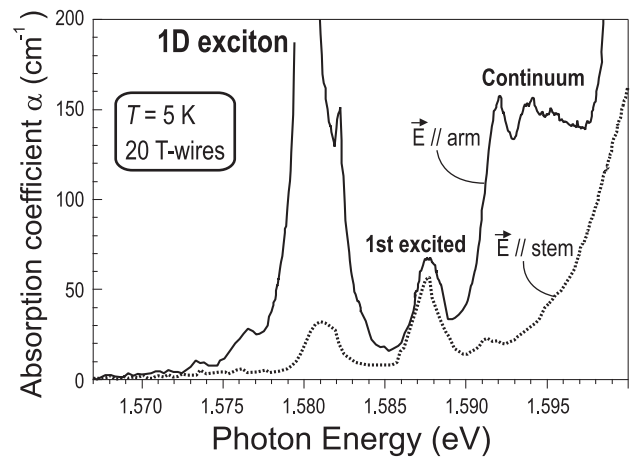}}}
\caption{Absorption spectra of 20 wires at 5 K. Solid curve and dotted curve show absorption spectrum for arm and stem polarizations, respectively.}
\label{fig3}
\end{figure}

The absorption spectrum for stem polarization has no peak structure. The difference in the absorption coefficient between the two polarizations is 145 cm$^{-1}$ at 1.4884 eV, which corresponds to e$^{-\alpha L}$=6.0$\times 10^{-4}$. The stronger absorption for arm polarization indicates that the T-wires have polarization dependence similar to the arm well \cite{yamaguchi1994}. This characteristic originates from the strong quantum confinement in the [110] direction. 

These experimental results on 1D excitonic absorption support the potential applications of quantum wires to various optical devices like modulators, switches, and amplifiers. The observed quantities are useful in the practical design of such optical devices.

Figure \ref{fig3} plots the absorption spectra for the 20 wires at 5 K. 
The solid and dotted curves correspond to the absorption spectra for arm and stem polarizations.
The spectral shape of the absorption for arm polarization agrees well with the results of our previous PLE study \cite{itoh2003} and the absorption spectrum for a single-wire device \cite{takahashi2005}, which assigns each peak to the 1D-exciton ground state, the first excited state of the 1D exciton, and the 1D continuum states, as denoted in the figure. The increasing absorption above 1.598 eV is due to exciton in the arm well.

The polarization dependence for each absorption peak is clearly observed. The absorption of the 1D exciton and continuum states decrease drastically in stem polarization. The excited exciton, on the other hand, has a similar absorption value for both polarizations. This difference originates from the difference in the hole subbands contributing to optical transition\cite{itoh2003}.

In this experimental setup, we were not able to measure absorption coefficients above 200 cm$^{-1}$ because of the limited transmission light. Although the peak value of the 1D exciton for arm polarization is above the detection limit, the FWHM is estimated to be 2 meV or less from that in stem polarization or that of the excited exciton state. This indicates a small amount of inhomogeneous broadening, or high uniformity in the 20 T-wires throughout the whole region. Therefore, the 7.2 meV FWHM at 297 K for the 1D exciton represents the intrinsic homogeneous width of the 20 T-wires.

We experimentally confirmed that the 1D-exciton peak at 5 K in Fig. \ref{fig3} changes continuously to the 1D-exciton peak at 297 K in Fig. \ref{fig2} as we increase temperature. 
The energy difference in the peaks between 297 and 5 K is 92.4 meV, which is close to the band-gap energy shift of 95 meV for bulk GaAs between the two temperatures \cite{chow1999}. 
These facts prove that the absorption peak at 297 K in Fig. \ref{fig2} is due to the 1D-exciton ground state. 
It should be stressed that this experiment demonstrates room-temperature 1D excitonic absorption, which has never been observed before. 

The lower columns in table \ref{table1} summarize the measured absorption data for a 20-wire device at 297 K and 5 K, and for a single-wire device at 5 K for arm polarization \cite{takahashi2005}. 

The absorption coefficient of the continuum states for arm polarization at 5 K is about 150 cm$^{-1}$. This value is 9.4 times greater than that for the single-wire device, 16 cm$^{-1}$, whose ratio corresponds to that for optical confinement factor $\Gamma$. 
This demonstrates that the so-called material absorption coefficient defined by $\alpha/\Gamma$ is the same in both 20-wire and single-wire devices for 1D continuum states at 5 K. 

The 1D-exciton peak area at 297 K is evaluated to be 1152 cm$^{-1}$ meV from the product of its peak value, 160 cm$^{-1}$, and the FWHM, 7.2 meV. We are not able to compare it with that for the 20-wire device at 5 K, because the peak absorption at 5K could not be measured. Instead, we compare it with the value of 128 cm$^{-1}$ meV for a single-wire device at 5K. The 1D-exciton peak area for the 20-wire device at 297 K is 9.0 times larger than that for a single-wire device at 5 K. Thus, the area of material absorption coefficient $\alpha/\Gamma$ is almost the same at both temperatures. This is reasonable because theories have predicted that the absorption area of the exciton peak should be preserved against changes in temperature \cite{chow1999,Haug2004}.  

In summary, we observed room-temperature 1D excitonic absorption in 20 T-wires embedded in an optical waveguide using waveguide-transmission spectroscopy. The absorption peak has a maximum value of 160 cm$^{-1}$ and a FWHM of 7.2 meV with strong polarization anisotropy, which support the application of quantum wires to practical optical devices. 

This work was partly supported by a Grant-in-Aid from the Ministry of Education, Culture, Sports, Science, and Technology, Japan.


\clearpage


\begin{references}
\bibitem{arakawa1982}Y. Arakawa and H. Sakaki, Appl. Phys. Lett. {\bf 40}, 939 (1982).

\bibitem{asada1985}M. Asada, Y. Miyamoto, and Y. Suematsu, Jpn. J. Appl. Phys., Part 2, {\bf 24}, L95 (1985).

\bibitem{akiyamaPL} H. Akiyama, J. Phys: Condens. Matter {\bf 10}, 3095 (1998), and references therein. 

\bibitem{itoh2003} H. Itoh, Y. Hayamizu, M. Yoshita, H. Akiyama, L. Pfeiffer, K. West, M. Szymanska, and P. Littlewood, Appl. Phys. Lett. {\bf 83}, 2043 (2003); H. Akiyama, M. Yoshita, L. N. Pfeiffer, A. Pinczuk, and K. W. West, Appl. Phys. Lett. {\bf 82}, 379 (2003). 

\bibitem{kapon1989} E. Kapon, D. Hwang, and R. Bhat, Phys. Rev. Lett. {\bf 63}, 430 (1989).

\bibitem{weg1993} W. Wegscheider, L. Pfeiffer, M. Dignam, A. Pinczuk, K. West, S. McCall, and R. Hull, Phys. Rev. Lett. {\bf 71}, 4071 (1993).

\bibitem{hayamizu2002} Y. Hayamizu, M. Yoshita, S. Watanabe, H. Akiyama, L. Pfeiffer, and K. West, Appl. Phys. Lett. {\bf81}, 4937 (2002)

\bibitem{yagi2003} H. Yagi, T. Sano, K. Ohira, T. Maruyama, A. Haque, and S. Arai, Jpn. J. Appl. Phys., Part 2, {\bf42}, L748 (2003)

\bibitem{takahashi2005} Y. Takahashi, Y. Hayamizu, H. Itoh, M. Yoshita, H. Akiyama, L. Pfeiffer, and K. West, Appl. Phys. Lett. {\bf 86}, 243101 (2005). 

\bibitem{pfeiffer1990} L. Pfeiffer, K. West, H. St\"ormer, J. Eisenstein, K. Baldwin, D. Gershoni, and J. Spector, Appl. Phys. Lett. {\bf 56}, 1697 (1990). 

\bibitem{yoshita2002} M. Yoshita, H. Akiyama, L. Pfeiffer, and K. West, Appl. Phys. Lett. {\bf 81}, 49 (2002).

\bibitem{Akiyama2003} H. Akiyama, L. N. Pfeiffer, M. Yoshita, A. Pinczuk, P. B. Littlewood, K. W. West, M. J. Matthews, and J. Wynn, Phys. Rev. B {\bf 67}, 041302(R) (2003).

\bibitem{takahashi2003} Y. Takahashi, S. Watanabe, M. Yoshita, H. Itoh, Y. Hayamizu, H. Akiyama, L. Pfeiffer, and K. West, Appl. Phys. Lett. {\bf 83}, 4089 (2003).

\bibitem{yamaguchi1994} A. Yamaguchi, K. Nishi, and A. Usui, Jpn. J. Appl. Phys. {\bf 33}, L912 (1994).

\bibitem{chow1999} W. W. Chow and S. W. Koch, {\it Semiconductor-Laser Fundamentals} (Springer, Berlin, 1999).

\bibitem{Haug2004} H. Haug and S. W. Koch, {\it Qnantum theory of the optical and electronic properties of semiconductors} (4th edition, World Scientific, Singapore, 2004). 

\end{references}
\end{document}